\documentclass[preprint]{revtex4}
\usepackage{graphicx}% Include figure files
\usepackage{dcolumn}% Align table columns on decimal point
\usepackage{bm}% bold math

\begin{document}

\title{Scaling of the magnetic entropy and magnetization in YbRh$_2$(Si$_{0.95}$Ge$_{0.05}$)$_2$\footnote{Manuscript submitted SCES2004 conference}}

\author{P. Gegenwart$^{(1)}$, Y. Tokiwa$^{(1)}$, K. Neumaier$^{(2)}$, C. Geibel$^{(1)}$, F. Steglich$^{(1)}$}
\address{$^{(1)}$ Max-Planck Institute for Chemical Physics
of Solids, D-01187 Dresden, Germany
\\ $^{(2)}$Walther Meissner Institute, D-85748 Garching, Germany}

\begin{abstract}

The magnetic entropy of YbRh$_2$(Si$_{0.95}$Ge$_{0.05}$)$_2$ is
derived from low-temperature ($T\geq 18$~mK) specific heat
measurements. Upon field-tuning the system to its
antiferromagnetic quantum critical point unique temperature over
magnetic field scaling is observed indicating the disintegration
of heavy quasiparticles. The field dependence of the entropy
equals the temperature dependence of the dc-magnetization as
expected from the Maxwell relation. This proves that the
quantum-critical fluctuations affect the thermal and magnetic
properties in a consistent way.

\end{abstract}

\maketitle

The heavy fermion system YbRh$_2$(Si$_{0.95}$Ge$_{0.05}$)$_2$ is
located very close to an antiferromagnetic (AF) quantum critical
point (QCP) \cite{Custers}. A small critical magnetic field of
$B_c\approx 0.027$~T is sufficient to suppress very weak AF order
from $T_N=20$~mK ($B=0$)  towards zero temperature. At magnetic
fields $b>0$, where $b=B-B_c$ denotes the difference between the
applied and critical field, a heavy Landau Fermi liquid (LFL)
state is induced below a characteristic temperature
$T_0(b)=1.09b$[Kelvin/Tesla] that increases linearly with $b$.
This cross-over between non-Fermi liquid and LFL behavior is
accompanied by unique $T/b$ scaling in thermodynamic and
transport properties observed over nearly four decades in
temperature over magnetic field \cite{Custers}. It indicates that
the characteristic energy of the heavy quasiparticles is governed
only by the ratio of the thermal energy to the magnetic field
increment $b$ and vanishes upon approaching the QCP. Such a
behavior is consistent with the {\it locally-critical} scenario
for an AF QCP, at which the heavy quasiparticles disintegrate
\cite{Schroeder,Si}.

In this paper, we derive scaling expressions for the magnetic
entropy and magnetization valid both in the LFL and non-Fermi
liquid region of the $B$-$T$ phase diagram. We prove that both
properties are fully consistent with each other and probe the
same degrees of freedom related to the QCP.

Specific heat data for 18~mK~$\leq~T\leq$~2~K and $0\leq~B\leq$~
0.8~T \cite{Custers} are used to calculate the magnetic entropy
$S(T,B)=\int_0^T C(T',B)/T'dT'$. At magnetic fields $B\geq
0.05$~T, for which a LFL state is well established at the lowest
measured temperature, the specific heat has been extrapolated
towards $T\rightarrow 0$, using $C=\gamma_0(B)T$ \cite{Custers}.
At smaller magnetic fields a constant specific heat coefficient is
not yet reached above 18~mK. Since the different entropy curves
at low magnetic fields $B\leq 0.1$~T merge above 1~K, the unknown
entropy contribution from the temperature interval $0\leq T\leq
18$~mK can be deduced with satisfactory precision (cf. error bars
in the inset of Fig. 1).

Fig. 1 displays the temperature dependence of the magnetic
entropy at different applied magnetic fields. At zero field, the
entropy gain at the AF phase transition amounts to only
$S(T_N)\approx 0.008R\log2$, indicating extremely weak AF order.
For undoped YbRh$_2$Si$_2$ the entropy at $T_N=70$~mK equals
$0.03R\log2$ \cite{Gegenwart Acta}. Thus, the ratio between the
ordering temperature and $S(T_N)$ remains unchanged.

We now turn to the unique $T/b$ scaling that, as discussed in the
introduction, hints at a locally-critical QCP in
YbRh$_2$(Si$_{0.95}$Ge$_{0.05}$)$_2$. Previously, the scaling
analysis of the specific heat has revealed
$C(T,b)/T=b^{-1/3}\Phi(T/T_0(b))$ with
$\Phi(x)\approx(\max(x,1))^{-1/3}$ \cite{Custers}. This implies
that the specific heat coefficient in the LFL state at $T\ll
T_0(b)$ diverges as $\gamma_0(b)\propto b^{-1/3}$. Such a stronger
than logarithmic mass divergence is clearly incompatible with the
predictions of the Hertz-Millis {\it itinerant} scenario
\cite{Hertz}. The corresponding scaling behavior of the magnetic
entropy is shown in Fig. 2a. Note that in the non-Fermi liquid
regime $T\gg T_0(b)$, the entropy is nearly magnetic field
independent and varies roughly as $S\propto T^{2/3}$. This
corresponds to $C/T\propto T^{-1/3}$, observed very close to the
critical field ($b\approx 0$) at temperatures below 0.4~K
\cite{Custers}.

Fig. 2b proves that the derivative $\zeta=\partial S/\partial B$,
experimentally deduced from the differential quotient of the
entropy data at different magnetic fields, equals the temperature
derivative $\partial M/\partial T$ of isofield dc-magnetization
measurements \cite{Tokiwa}, as expected from the Maxwell relation.
Both thermal and magnetic properties are thus influenced by the
nearby QCP in a consistent way. In the LFL state ($T\ll T_0(b)$)
$\zeta\propto-Tb^{-4/3}$ and the magnetic susceptibility can be
derived using $\chi={\partial M\over\partial
B}={\partial\over\partial
B}\int_0^T\zeta(T',b)dT'=\chi_0(b)+T^2b^{-7/3}$. Thus, $\chi$
approaches the Pauli susceptibility $\chi_0(b)={\partial
M\over\partial B}|_{T=0}$ with a $T^2$ temperature dependence and
the coefficient of this term diverges strongly upon approaching
the QCP at $b\rightarrow 0$. Such behavior has been observed in
ac-susceptibiliy measurements on undoped YbRh$_2$Si$_2$
\cite{Trovarelli Letter}.

%The scaling analysis also provides information on the temperature
%dependent part $\Delta\chi=\chi(T,b)-\chi_0(b)$ of the magnetic
%susceptibility in the LFL regime: We obtain $\Delta\chi\propto
%T^2b^{-7/3}$. Interestingly, the field dependence of this $T^2$
%contribution to the magnetic susceptibility is different from
%that of the Pauli susceptibility $\chi_0(b)$ \cite{Bemerkung}.

%indicating that the free energy must contain a term that is field
%but not temperature dependent.

To summarize, we have analyzed the magnetic entropy of
YbRh$_2$(Si$_{0.95}$Ge$_{0.05}$)$_2$. At zero magnetic field, a
tiny entropy of about 0.8\%$R\log 2$ is related to the
antiferromagnetic state. In the field-induced LFL state, the
entropy scales as $S(T,b)\propto Tb^{-1/3}$ and its magnetic field
dependence is consistent with the temperature dependence of the
magnetization. This proves that the specific heat and the
temperature dependent part of the magnetic susceptibility probe
the same degrees of freedom, related to the nearby
locally-critical QCP.

%The temperature dependent part of the magnetic susceptibility in
%the LFL regime scales as $\Delta\chi\propto T^2b^{-7/3}$.

We gratefully acknowledge discussions with C. P\'{e}pin and I.
Paul.

%----------------------------------------------------------------------
% Reference section
%
% List each reference with a separate \bibitem{} command.  The
% argument contains the label that is used in the \cite{} command
% in the main text
%
% e.g.
%
%    This follows our pioneering work on TdB2\cite{TdB2}.
%
% \bibitem{TdB2}
% J. Doe, J. Doe, and J. Dupont, J. Irrep. Res. 10 (2000) 1000.

\newpage
\begin{figure}
\centerline{\includegraphics{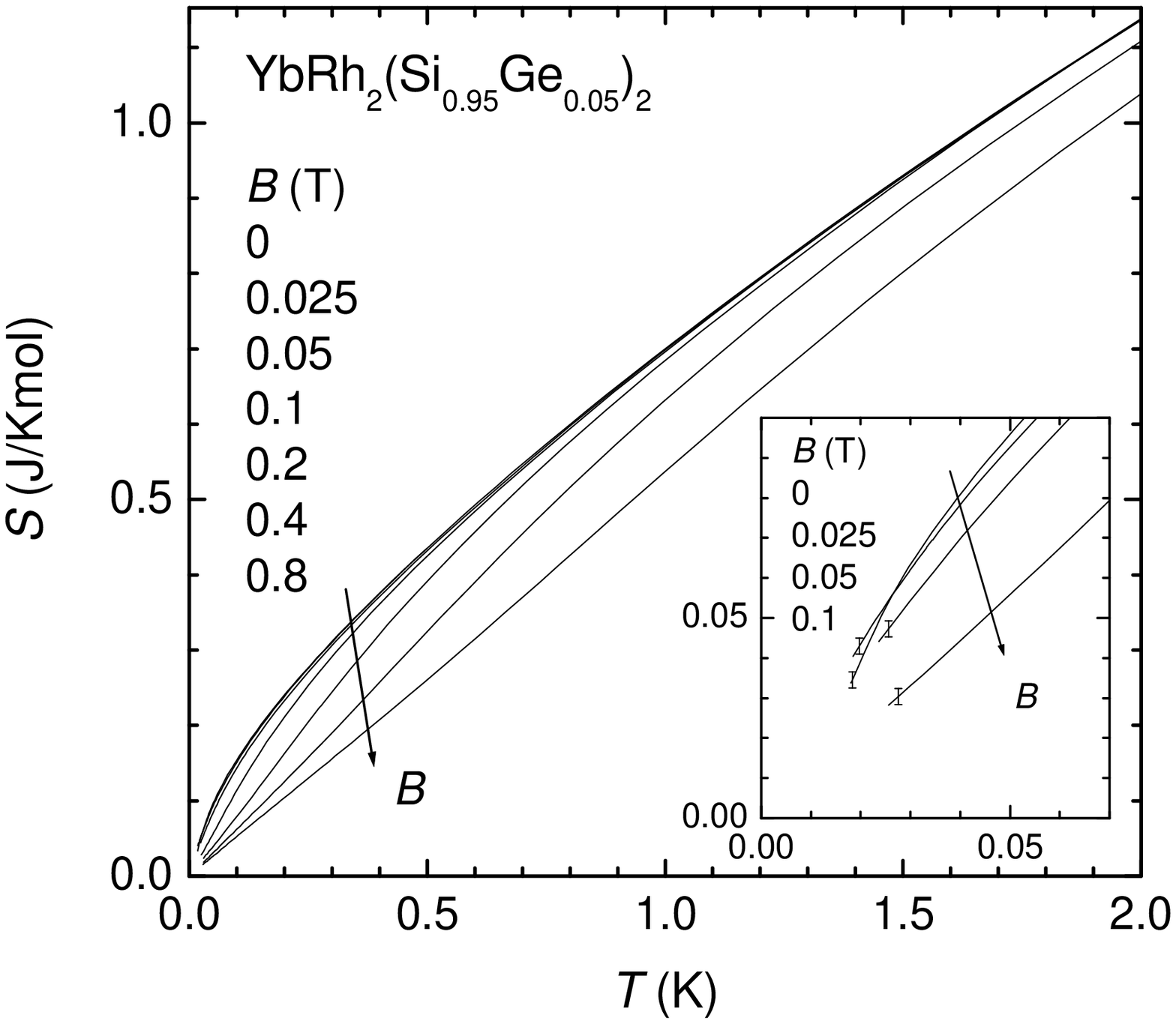}}
\caption{Magnetic entropy $S(T)$ of
YbRh$_2$(Si$_{0.95}$Ge$_{0.05}$)$_2$ at various magnetic fields,
obtained from integration of specific heat measurements
\cite{Custers}. Inset enlarges low-$T$ behavior.} \label{Fig1}
\end{figure}

\begin{figure}
\centerline{\includegraphics{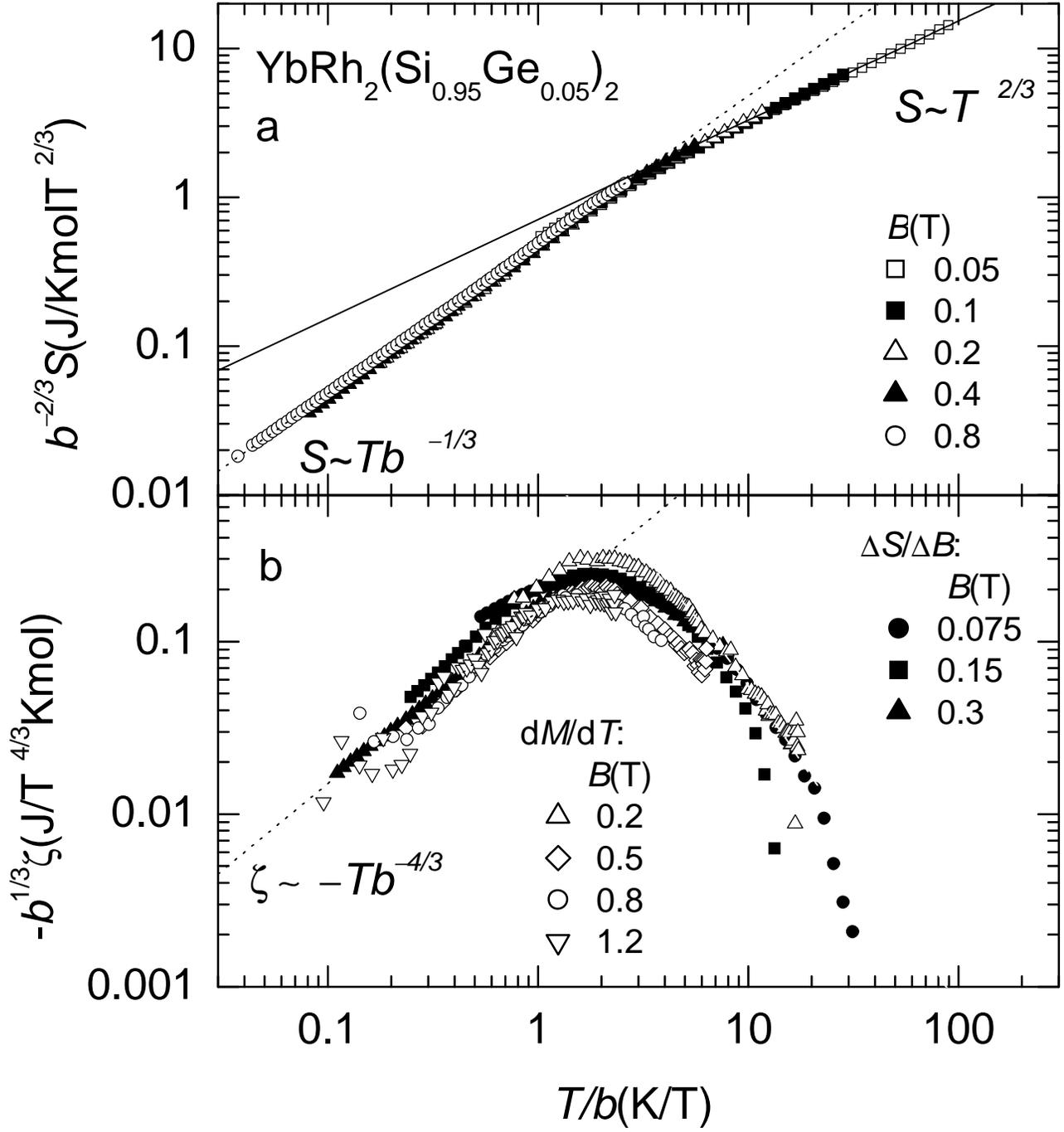}} \caption{$T/b$
scaling for YbRh$_2$(Si$_{0.95}$Ge$_{0.05}$)$_2$. Entropy $S$ as
$b^{-2/3}S$ vs $T/b$ on log-log plot (a). Dashed (dotted) line
represents $S=0.48$~JT$^{1/3}$K$^{-2}$mol$^{-1}\times Tb^{-1/3}$
($S=0.71$~JK$^{-5/3}$mol$^{-1}\times T^{2/3}$). Second derivative
of free Energy $\zeta=\partial^2F/\partial B\partial T=\partial
S/\partial B=\partial M/\partial T$ as $-b^{1/3}\zeta$ vs $T/b$
on log-log plot (b). Open and closed symbols represent values
obtained from differential quotient $\Delta S/\Delta B$ and slope
$dM(T)/dT$ of isofield dc-magnetization measurements
\cite{Tokiwa}, respectively. Dotted line represents
$\zeta=-0.15$~T$^{1/3}$JK$^{-2}$mol$^{-1}\times Tb^{-4/3}$.}
\label{Fig2}
\end{figure}

%----------------------------------------------------------------------
% Terminate document

\end{document}